\documentclass[%
 aip,
 amsmath,amssymb,
 reprint,%
]{revtex4-1}

\pdfoutput=1
\usepackage{graphicx}% Include figure files
\usepackage{dcolumn}% Align table columns on decimal point
\usepackage{bm}% bold math

\usepackage{xcolor}
\usepackage[normalem]{ulem}

\usepackage[utf8]{inputenc}
\usepackage[T1]{fontenc}
\usepackage{mathptmx}

\begin{document}

\title[]{Thomson scattering on the Large Plasma Device}
% Force line breaks with \\

\author{S. Ghazaryan}
\thanks{These two authors contributed equally}
%\email{sofighazaryan82@g.ucla.edu.}
%\affiliation{Department of Physics and Astronomy, University of California Los Angeles, Los Angeles, CA 90095, USA}
\affiliation{Department of Physics and Astronomy, University of California Los Angeles, Los Angeles, CA 90095, USA}

\author{M. Kaloyan}
\thanks{These two authors contributed equally}
\affiliation{Department of Physics and Astronomy, University of California Los Angeles, Los Angeles, CA 90095, USA}

\author{W. Gekelman}
\affiliation{Department of Physics and Astronomy, University of California Los Angeles, Los Angeles, CA 90095, USA}

\author{Z. Lucky}
\affiliation{Department of Physics and Astronomy, University of California Los Angeles, Los Angeles, CA 90095, USA}

\author{S. Vincena}
\affiliation{Department of Physics and Astronomy, University of California Los Angeles, Los Angeles, CA 90095, USA}

\author{S.K.P. Tripathi}
\affiliation{Department of Physics and Astronomy, University of California Los Angeles, Los Angeles, CA 90095, USA}

\author{P. Pribyl}
\affiliation{Department of Physics and Astronomy, University of California Los Angeles, Los Angeles, CA 90095, USA}

\author{C. Niemann}
\affiliation{Department of Physics and Astronomy, University of California Los Angeles, Los Angeles, CA 90095, USA}

%\date{\today}% It is always \today, today,
             %  but any date may be explicitly specified

\begin{abstract}
We have developed a non-collective Thomson scattering diagnostic for measurements of electron density and temperature on the Large Plasma Device.
A triple grating spectrometer with a tunable notch filter is used to discriminate the faint scattering signal from the  stray light. In this paper, we describe the diagnostic and its calibration via Raman scattering, and present the first measurements performed with the fully-commissioned system. Depending on the discharge conditions, the measured densities and temperatures range from 4.0$\times$10$^{12}$ cm$^{-3}$  to 2.8$\times$10$^{13}$ cm$^{-3}$, and from 1.2~eV to 6.8~eV, respectively. The variation of the measurement error with plasma parameters and discharges averaged is also discussed. 
\end{abstract}

\maketitle

\section{\label{sec:introduction}Introduction}
Thomson scattering (TS) is a powerful, first-principles, and non-invasive plasma diagnostic that derives electron density and temperature from the Doppler shift imparted to photons scattered by plasma electrons \cite{evans}.
In the non-collective regime encountered in tenuous basic plasmas, the temperature can be derived from the width of the broadened line, and the density from the scattering signal intensity, after an absolute irradiance calibration. 
%
%TS measurements in this regime are more challenging than in higher density and temperature fusion plasmas \cite{froula12}, 
TS measurements in this regime are challenging 
%since the spectral width is small and 
since the faint scattering signal must be discriminated from a much brighter background laser light.
However, TS spectra from such plasmas can be resolved by using a suitable notch filter for stray light suppression \cite{muraoka1998, kono2000,vds, yamamoto12, lee2018, vincent2018,Shi2021}.
% %zhao2019, belostotskiy2008

We have developed a TS system on the Large Plasma Device (LAPD), a 22~m long magnetized linear plasma device at the University of California Los Angeles \cite{gekelman2016}. 
The instrument allows 
model-independent measurements of the density and temperature in experiments involving fast laser-driven flows at ion-scales \cite{niemann2014,schaeffer2022}, that cannot be diagnosed with swept electrostatic probes \cite{chen}.
Scattering spectra can be obtained by accumulating the signal from as little as ten discharges.%, and with high spatial (mm) and temporal (4~ns) resolution.

\section{\label{sec:setup}Experimental setup}
\begin{figure}[!ht]
    \centering 
    \includegraphics[width=\linewidth]{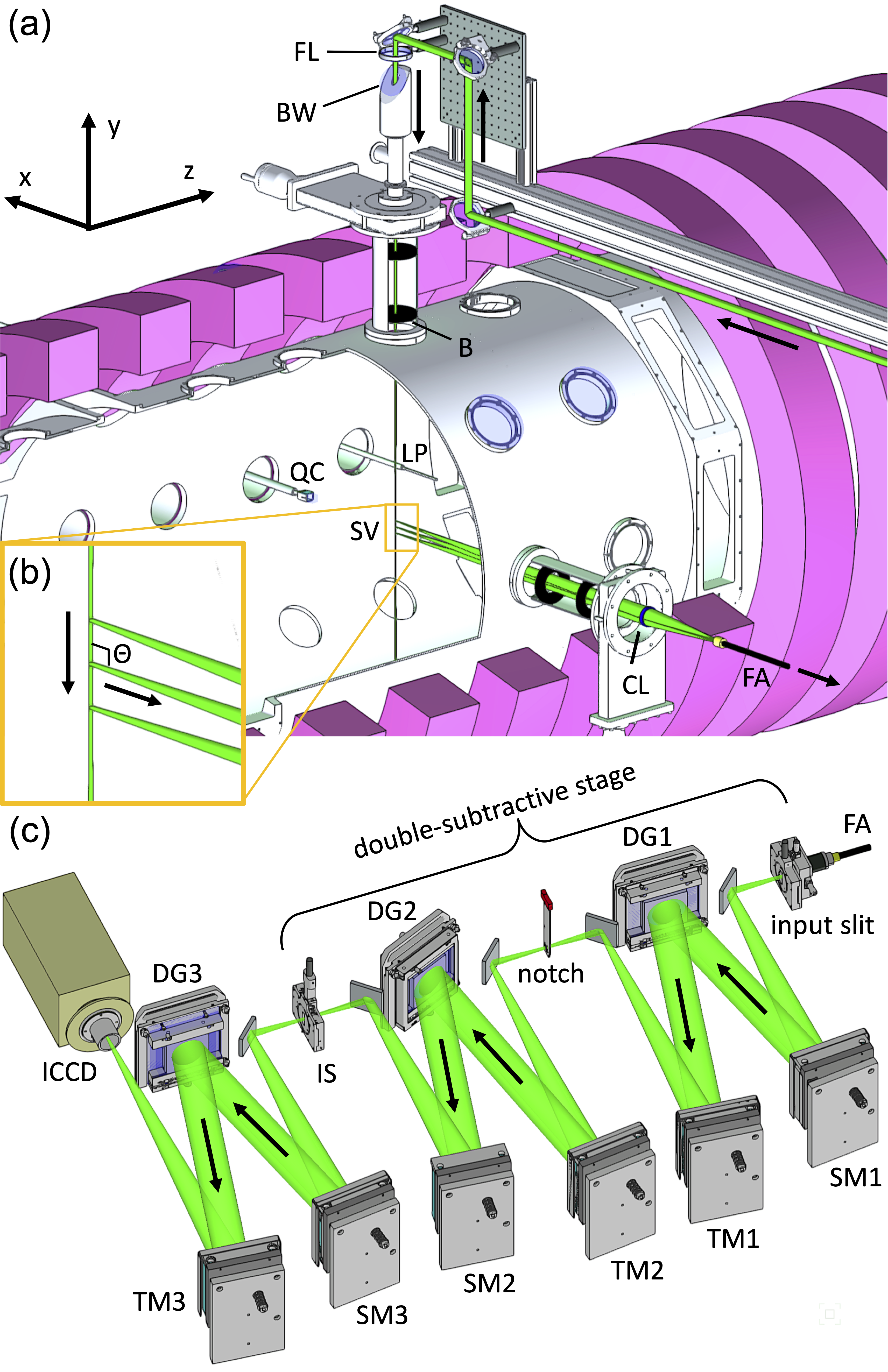}
    \caption{(a) Schematic of the setup on the LAPD. The laser beam is focused into the scattering volume (SV) by a focusing lens (FL) and enters and exits the chamber top to bottom through Brewster windows (BW) and a series of baffles (B). The scattering volume is imaged onto the  fiber array (FA) by the collection lens (CL). 
    The chamber and pink magnets have been cut to provide a view on the inside. 
    Also shown are the Langmuir probe (LP), and the quartz crystal (QC) used for alignment and calibration.
    (b) Scattering volume showing light collected by the top, center, and bottom fiber at an angle $\Theta=90^o$. (c) Schematic of the light path through the spectrometer showing the three diffraction gratings (DG), notch mask, intermediate slit, and the spherical (SM) and toroidal (TM) mirrors.
    }
    \label{fig:setup}
\end{figure}

A schematic of the setup is shown in Fig. \ref{fig:setup}a.
Large plasmas (19~m length and 50~cm diameter) are created by accelerating electrons emitted from a heated cathode into a gas by an electrostatic field.
In this experiment, the plasmas are magnetized by a straight axial magnetic field of 1.0 kG, and pulsed on for 15~ms at 1/3~Hz repetition rate. Helium gas is injected near the plasma source by two fast Piezo valves pulsed on 5~ms prior to and for the duration of the discharge (80~V gate voltage, 450 SCCM, 33 PSI backing pressure). 
%The plasma expands into vacuum as it outruns the gas.
%
The TS diagnostic is installed in port 32, about 13 m from the plasma source.

A frequency-doubled Nd:YAG laser produces 4~ns long pulses of 460~mJ energy at a wavelength of $\lambda_i$=532~nm, at a repetition rate of 20~Hz.
The laser output is collimated to a 15~mm diameter beam using a Galilean telescope and is focused by a 1.5~m focal-length lens into the center of the plasma column. This small-solid angle f/67 lens produces a cylindrical beam around the focus with an  0.8~mm full-width at half-maximum (FWHM) over a length of several centimeters.
The beam enters and exits the chamber from top to bottom through Brewster windows and a series of baffles, that minimize stray light entering the detector. 
A half-waveplate linearly polarizes the incident beam along the longitudinal axis ($\hat{z}$). 
The laser energy was monitored on each shot using a pyroelectric sensor and was found to be stable within $\pm 2.5\%$.

Scattered light is collected perpendicularly to the injection path ($\Theta$=90$^o$) by a 50~mm diameter and 20~cm focal-length lens, located at a distance of 94~cm from the beam focus. This lens projects a 3.7$\times$ demagnified image of the beam onto a linear fiber bundle.
Using relatively slow focusing and collection lenses outside the vacuum chamber allows the diagnostic to move from port to port and minimizes alignment sensitivity, and yet results in high light-collection efficiency with an effective optical extent of 0.9$\times G_{max}$, where $G_{max}$ is the spectrometer extent \cite{kaloyan2022}.
% ######
The fiber array consists of 40 linearly arranged 200~$\mu$m diameter core optical fibers with a numerical aperture of 0.22, mapped end-to-end with combined slit dimensions of 8.8~mm $\times$ 0.22~mm.
The intersection of the beam and fiber array projection defines the scattering volume (Fig. \ref{fig:setup}b). Integrating light from all 40 fibers results in a 33~mm long $\times$ 0.74 mm wide vertical scattering volume, while the spatial resolution of a single fiber is sub-millimeter. The 25~m fibers are coupled directly to the input of the spectrometer.

% #######
A custom triple grating f/4 Czerny-Turner imaging spectrometer
is used to resolve the scattering signal (Fig. \ref{fig:setup}c).
Light from the fibers passes through three monochromator stages before reaching the detector. 
%Light from the input slit is collimated by the spherical input mirror (SM1), diffracted by the first grating (DG1), and condensed and focused by the toroidal exit mirror (TM1) onto the notch mask. 
%
The first stage images the spectrum onto a mask, where a 0.75~mm wide 
%and 50~$\mu$m thick 
stainless steel notch blocks a 1.5~nm range around $\lambda_i$.
Dispersion caused by the first stage is canceled by the second stage. 
This allows the double-subtractive system to operate as a tunable notch filter.
The final stage disperses the stray light subtracted spectrum onto the detector.
Holographic gratings, in combination with baffles, and an intermediate slit between the second and third stages are used to minimize stray light.
The three 110~mm $\times$ 110~mm aluminium-coated,  1200 grooves/mm gratings are blazed at 500 nm, for an efficiency around 60$\%$ at 532~nm. All other reflectors are silver-coated to maximize throughput.
The measured transmission without the notch is around 20$\%$ at $\lambda_i$. 
Input and output focal lengths of each stage are 50 cm and 55 cm, respectively. 
Using an asymmetric system with two different focal lengths permits high throughput and spectral resolution, while maintaining off-axis angles large enough to avoid vignetting. %, and minimizing stray light.
Toroidal mirrors provide compensation for inherent astigmatism introduced by off-axis spherical reflectors. 
%The use of asymmetric design in conjunction with high-blaze gratings balances the tradeoffs between high luminosity and high notch extinction ratio over the length of the slit and fiber array.
%
%
%Toroidal,  102~mm $\times$ 102~mm mirrors are used in each stage to compensate for the inherent astigmatism introduced by off-axis spherical reflectors. %\cite{shafer64}.
%
The system was designed with this tunable notch instead of a volume Bragg grating \cite{vincent2018} to be compatible with 527~nm and experiments that require higher laser energies \cite{schaeffer2016}.

Spectra are recorded on an image intensified charge couple device (ICCD) equipped with a generation III photocathode and a quantum efficiency of around 50$\%$ at 532~nm. The micro-channel plate is gated at 10~ns and at maximum gain. Despite using 2$\times$2 hardware binning the average pixel count is only a small fraction of the 16-bit maximum and well within the linear response range. For all data presented here, the spectra were also software binned over 512 vertical pixels (all 40 fibers) and over two pixels horizontally into 256 total bins with 0.0776~nm/bin. A background image recorded without the plasma is subtracted from each TS spectrum.

\begin{figure}[!b]
    \centering 
    \includegraphics[width=\linewidth]{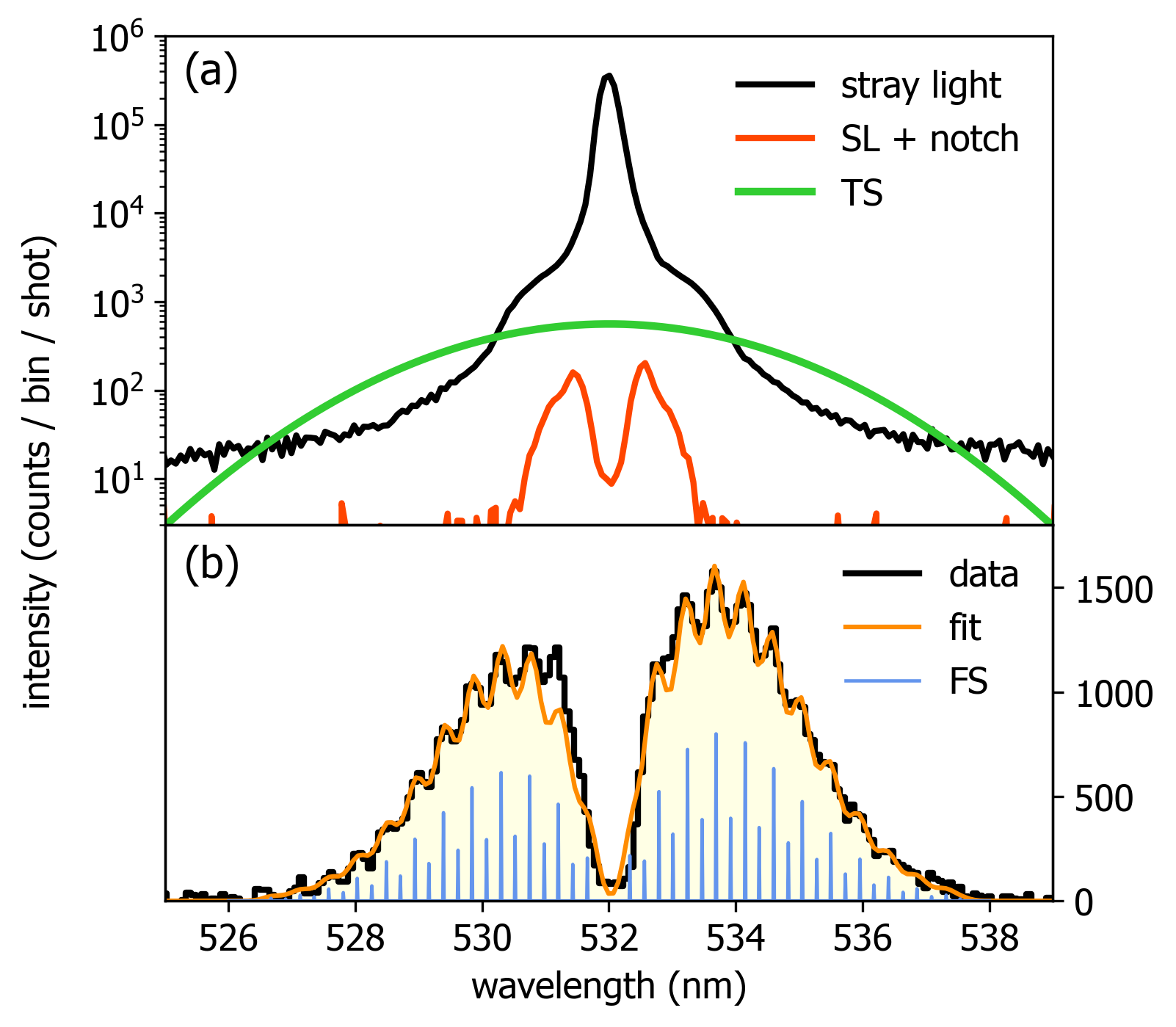}
    \caption{(a) Laser stray light is used to measure the instrument function (black). The profile shows a Gaussian peak with 0.28~nm FWHM and the broad Lorentzian wings caused by diffraction off of the gratings. A simulated TS spectrum for 1.1$\times$10$^{13}$ cm$^{-3}$ and 4.7~eV is shown for comparison (green). The notch reduces the stray light to well below the TS signal across the entire spectral range (red). (b) Measured Raman scattering spectrum from nitrogen at 10 torr (black) and theoretical fit for T=298~K (orange) used for the irradiance calibration. The fine structure (blue) due to the rovibrational lines is mostly washed out by the broad instrument function.}
    \label{fig:calibration}
\end{figure}
Figure \ref{fig:calibration}a shows the stray light spectrum measured without the notch and represents the 
instrument function. The instrument profile is the convolution of the Gaussian contribution due to the aberrated slit width, and the Lorentzian profile caused by diffraction off of the three gratings. 
Both the input and intermediate slits were fully open (4~mm) to maximize light throughput, and 
so the fiber array itself serves as 0.20~mm wide slit. 
This configuration provides a spectral resolution of 0.28~nm (FWHM) over a spectral range of 19.8~nm.
A calculated TS spectrum for 4.7~eV and 1.2$\times$10$^{13}$ cm$^{-3}$ is shown for comparison (from section \ref{sec:results}).
Without the notch, the stray light at $\lambda_i$ exceeds the TS signal by more than three orders of magnitude, and the broad wings drown out the signal even far from $\lambda_i$. While stray light can be subtracted using background spectra, its shot noise cannot. The notch filter reduces the stray light to well below the faint TS signal throughout the entire spectrum (red line) and makes its detection possible.
The notch reduces the intensity of the wings outside the 1.5~nm blocking range, since it removes stray-light before it can diffract and scatter off of the final two gratings. 

Raman scattering off of gas was used for 
an absolute irradiance calibration. For this purpose, the chamber was filled with nitrogen at (10.0 $\pm$ 0.1)~torr.
This pressure results in a Raman signal comparable to TS in amplitude and width, which minimizes potential inaccuracies related to a non-linear detector response. 
Figure \ref{fig:calibration}b shows the measured Raman signal obtained by averaging the scattered light from 5,000 laser shots.
The spectrum consists of dozens of peaks on either side of $\lambda_i$. Each line corresponds to a different rovibrational transition from one rotational state to another, induced by the inelastic scattering process \cite{quartz}. 
The fine structure (blue) is mostly washed out by the instrument function, which allows only to resolve the brightest lines.
A theoretical fit for a gas temperature of 298~K obtained as described in detail elsewhere \cite{vds} and convoluted with the instrument function
reproduces the measured spectrum well (orange), including the difference in brightness of the red-shifted Stokes and the blue-shifted anti-Stokes lines \cite{quartz}.  
Although the notch blocks most of the elastic Rayleigh scattering line that has an amplitude about 2,000 times that of the Raman signal, a small fraction leaks through the edge of the notch and contributes to the measured spectrum close to $\lambda_i$. Therefore, the calibration factor is determined from the area under the fit (shaded orange), which is slightly smaller than the integrated measured Raman signal. 
For nitrogen, the ratio between the TS signal (counts) and the Raman signal is $N_T/N_R = 1.23\times 10^4\cdot (n_e/n_{gas})$, where $n_e$ is the electron density responsible for TS, and $n_{gas}=3.24\times 10^{17}$ cm$^{-3}$ is the gas density responsible for the Raman signal \cite{vds}. In this experiment the electron density can, therefore, be determined from the measured TS signal $N_T$ as
\begin{equation}
    n_e = (2.98\pm 0.20)\times 10^8\ cm^{-3}\cdot N_T\ .
\end{equation}
This factor agrees within 10$\%$ with a redundant calibration obtained from Raman scattering off of a quartz crystal, that can be inserted into the scattering volume in lieu of the gas \cite{quartz}.

\section{\label{sec:results}Results and discussion}
When the scattering is non-collective, as it is in this experiment, the scale length of the electron density fluctuation sampled by TS ($\sim 1/k$) is small compared to the electron screening length $\lambda_D$. Here $k = 4\pi \cdot sin(\Theta/2)/\lambda_i$ is the scattering vector, and $\lambda_D = \sqrt{\epsilon_0 k_B T_e / n_e e^2}$ is the Debye length. The scattering parameter is then $\alpha = 1/k\lambda_D \ll 1$.
If the energy distribution is Maxwellian, the TS spectrum has a Gaussian shape, and the temperature can be determined from its width
\begin{equation}
T_e = \frac{m_ec^2}{8 k_B sin^2(\Theta/2)} \cdot \left( \frac{\Delta \lambda_{1/e}}{\lambda_i} \right)^2 .
\label{formula}
\end{equation}
Here $m_e$ is the electron mass, $c$ is the speed of light, $k_B$ is the Boltzmann constant,
and $\Delta\lambda_{1/e}$ is the spectral (e$^{-1}$) half-width. For $\Theta$=90$^o$ and $\lambda_i$=532~nm the formula can be simplified to $T_e$ (in eV) = 0.4513$\cdot (\Delta\lambda_{1/e})^2$.

\begin{figure}[!ht]
    \centering 
    \includegraphics[width=\linewidth]{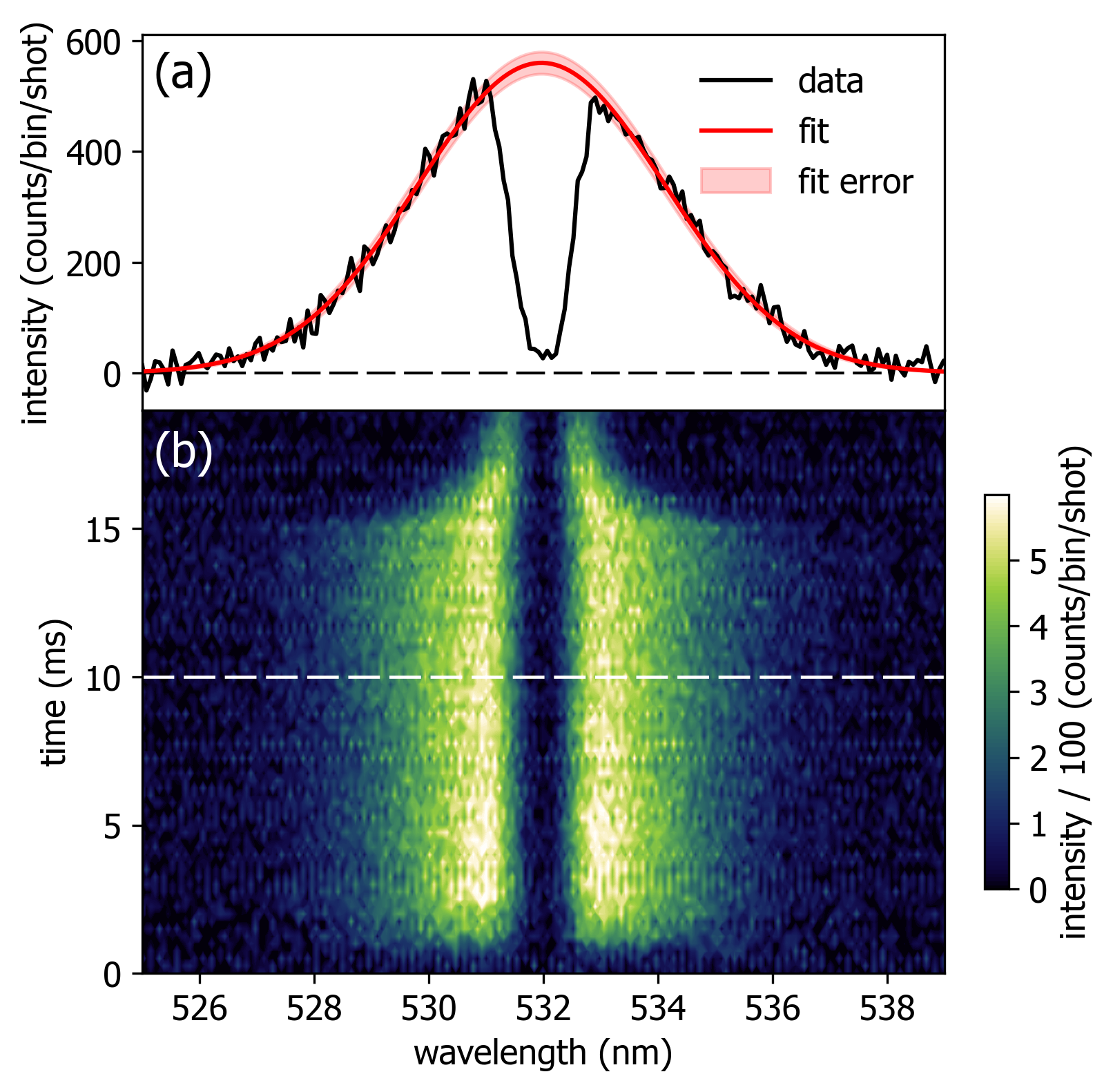}
    \caption{(a) The TS spectrum at t=10~ms after breakdown, obtained by integrating the signal from 2,000 discharges. 
    (b) Evolution of the TS spectrum during the 15~ms long discharge. The streak-plot combines 77 individual spectra recorded in steps of 0.25~ms.}
    \label{fig:streak}
\end{figure}
Figure \ref{fig:streak}a shows a TS spectrum from a plasma discharge in helium, recorded at 10~ms after the breakdown. 
The spectra from 2,000 discharges were averaged to
increase the signal-to-noise ratio (SNR).
The profile has a Gaussian shape, with the central 1.5~nm section suppressed by the notch. 
The Gauss-fit is consistent with $T_e = (4.7\pm 0.1)$ eV and $n_e= (1.2\pm 0.1)\times 10^{13}$ cm$^{-3}$.
The scattering parameter is $\alpha$ = 1.3$\times$ 10$^{-2} \ll$ 1. 
Figure \ref{fig:streak}b shows the evolution of the TS spectrum in time. This streak-plot is constructed from %77
individual spectra obtained in steps of 0.25~ms, with 200 shots averaged per time step. 
During these discharges, the spectral width slowly increases  with time, while the intensity decreases. When the discharge ends at 
t=15~ms the width quickly decreases. The evolution of $T_e$ and $n_e$ can be derived from this data and is compiled in figure \ref{fig:streak}, along with other discharge and plasma parameters.     
\begin{figure}[!ht]
    \centering 
    \includegraphics[width=\linewidth]{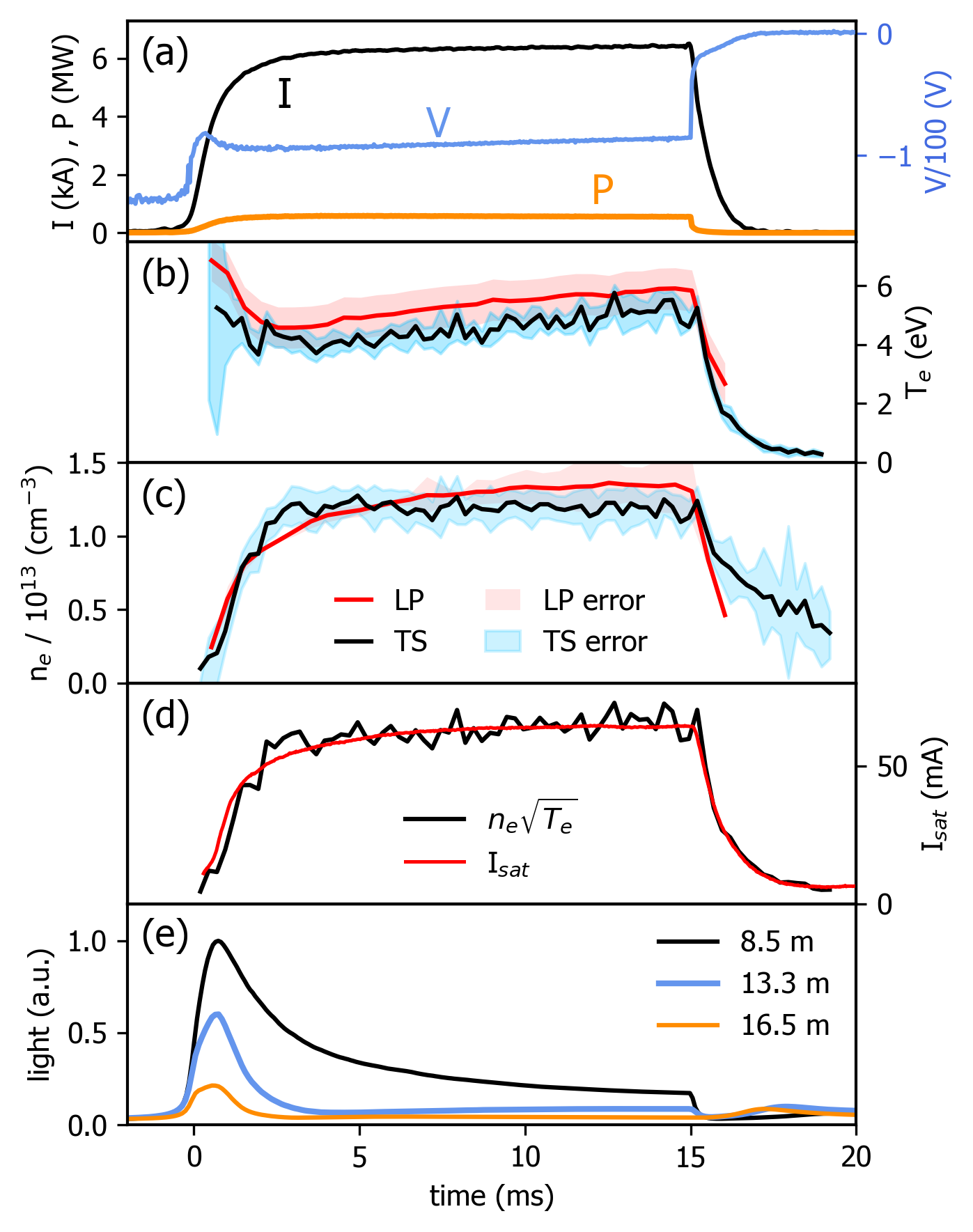}
    \caption{Evolution of select measured discharge and plasma-parameters as a function of time: (a) Discharge current, voltage, and power. (b) and (c) Electron temperature and density measured by TS and by a Langmuir probe (LP). (d) Ion saturation current from the Langmuir probe and comparison to $n_e\sqrt{T_e}$ from the TS data, scaled to fit the current. (e) Photodiode measurements of the plasma visible light self-emission at different distances from the source.}
    \label{fig:time}
\end{figure}
A capacitor bank voltage of 140~V results in an average discharge current of 6.3~kA and a power of 0.58~MW at this gas pressure. The time that the plasma current crosses a threshold value of about 1~kA defines t=0.
Within the error bars, the measured $T_e$ initially decreases from 5.5~eV after the breakdown to 4~eV around t=3~ms, before gradually increasing up to 5.2~eV. The measured density remains constant at 1.2$\times$10$^{13}$ cm$^{-3}$ throughout the discharge until t=15~ms, and then decreases approximately exponentially.   
TS measurements of $T_e$ agree well with data derived from the current-voltage (I-V) trace measured by a swept Langmuir probe \cite{chen}, that show a similar temporal evolution of $T_e$ but 20$\%$ higher temperatures. This discrepancy could be due to the fact that the probe was located 32~cm from the scattering volume and closer to the plasma source. The Langmuir probe is also sensitive to electrons in the tail of the distribution, while TS is less so.
%
%An accurate determination of $n_e$ from the Langmuir probe was not possible since the line-integrated microwave interferometer used to calibrate the relative density profile was two meters closer to the source than the probe.
TS measurements of $n_e$ also agree within 20$\%$ with Langmuir probe data.
Relative probe measurements of the electron density profile $n(x)$, are absolutely calibrated with the line-integrated density measured by a 100~GHz microwave interferometer, although in a port two meters closer to the source. 
Figure \ref{fig:time}d shows the ion-saturation current measured with the Langmuir probe biased negatively. The current increases steadily and peaks just before the discharge ends.
As expected \cite{chen}, the curve agrees well with $\sim n_e \sqrt{T_e}$ calculated from the TS data, and scaled to match the current. 
Plasma self-emission measurements performed with a silicon photodiode, 
%at different distances from the source, and
integrating all wavelengths between 200~nm and 1100~nm (Fig. \ref{fig:time}e), show a strong axial dependence of the intensity and temporal evolution on the distance from the source. 
This suggest an axial variation of $T_e$, consistent with Langmuir probe measurements in other experiments, that have observed much higher temperatures ($T_e >$10~eV) near the source.

\begin{figure}[!ht]
    \centering 
    \includegraphics[width=\linewidth]{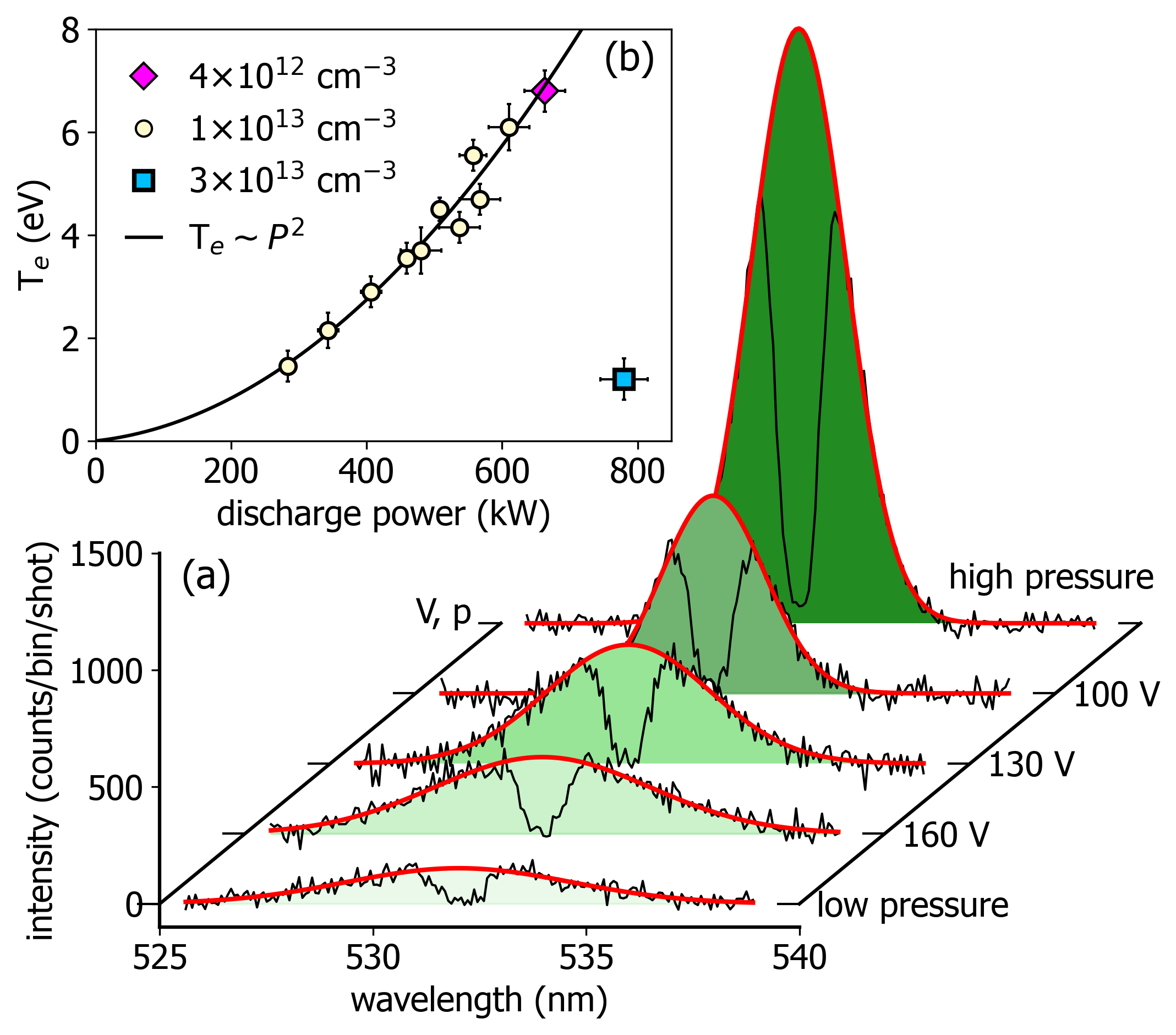}
    \caption{(a) TS spectra for various capacitor bank voltages and gas pressures. Increasing the bank voltage or lowing the gas pressure increases the temperature, as indicated by a broader and lower amplitude profile. (b) Measured $T_e$ as a function of discharge power for three different gas pressures and densities. The power was varied by changing the bank voltage from 100~V to 160~V. For a given pressure, the temperature increases with power as $T_e\sim P^2$.}
    \label{fig:power}
\end{figure}
Scattering spectra for different capacitor bank voltages and gas pressures measured at t=10~ms are shown in Fig. \ref{fig:power}. The temperature increases with the bank voltage and as the gas pressure is lowered, resulting in broader and smaller amplitude spectra. 
The temperature increases with discharge power $P$.
For the LAPD operating mode described here and for a constant inlet pressure, the trend is $T_e\sim P^2$ (Fig. \ref{fig:power}b). 
%
%For a constant gas pressure the measured electron temperature increases with power as $T_e\sim P^2$ (Fig. \ref{fig:power}b). 
%
A lower gas pressure and plasma density result in a higher temperature for a given bank voltage, in accordance with a simple discharge model \cite{lieberman}. Maximizing the gas pressure by pulsing the Piezo valves at maximum voltage ($\approx$1500 SCCM total flow) raises the plasma density to 2.8$\times$10$^{13}$ cm$^{-3}$ but leads to a much lower temperature $T_e=1.2$~eV. Near the plasma source the temperatures are likely much higher, with a large axial gradient in electron temperature in this particular operation (modest input power and a high feedstock gas fill rate).

\begin{figure}[!t]
    \centering 
    \includegraphics[width=\linewidth]{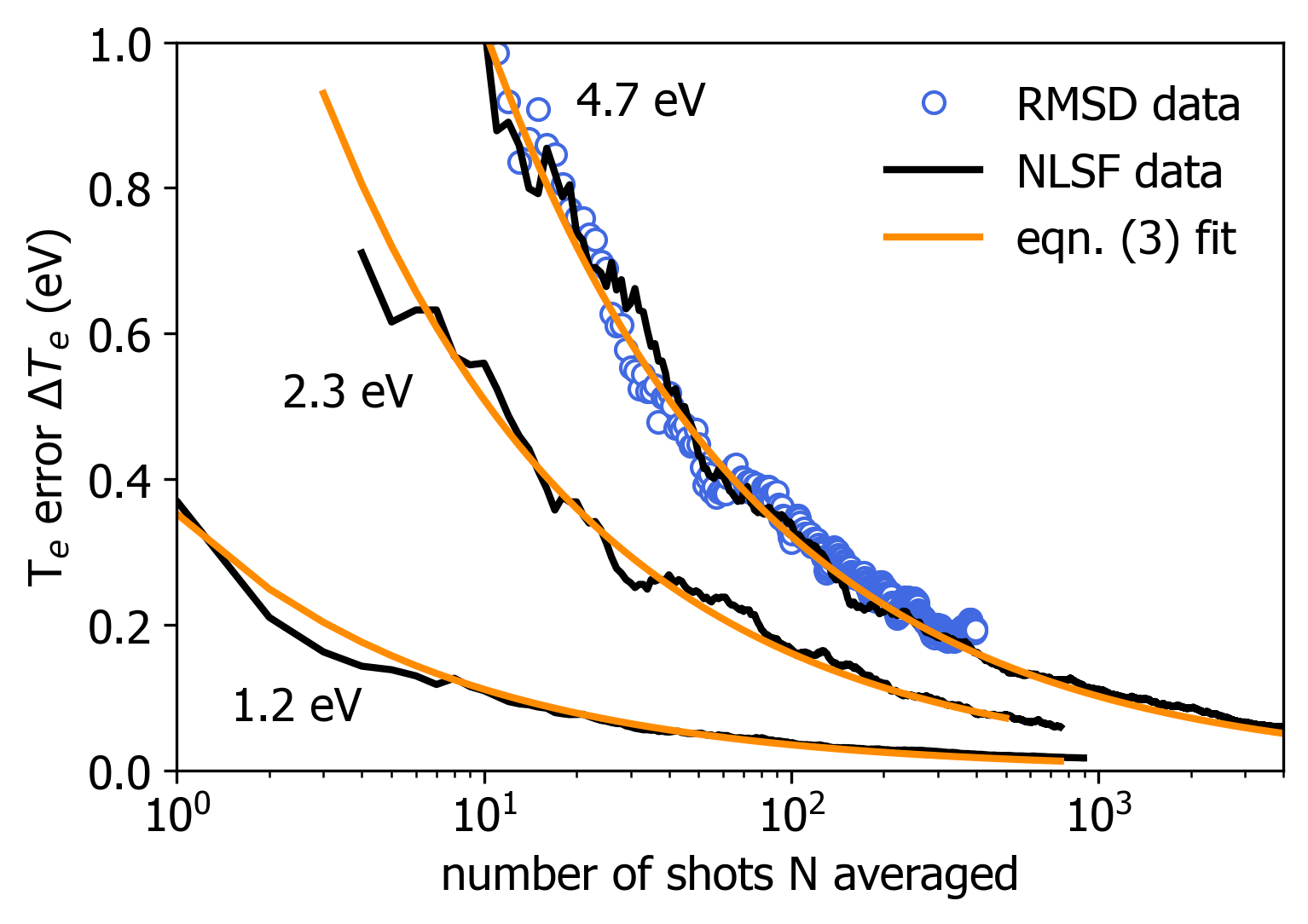}
    \caption{Measurement $T_e$ error as a function of discharges $N$ averaged, for three different temperatures. The data is consistent with equation 3 (orange curves).}
    \label{fig:error}
\end{figure}

The accuracy of the temperature measurement depends on the SNR.
Shot noise due to plasma self emission is negligible with the short exposure time.
The 20 e$^-$ / pixel readout noise of the cooled ICCD (-20$^o$C) is also negligible when binning and averaging multiple images, and so the total noise is determined solely by the shot noise of the TS signal. Based on Poisson statistics, the SNR increases with the number $N$ of spectra averaged as SNR$\sim \sqrt{N}$. 
Figure \ref{fig:error} shows the error $\Delta T_e$ measured for different temperatures as a function of $N$.  
The error is directly determined by the nonlinear least squares fit (NLSF) \cite{lmfit} used to fit the Gaussian spectrum.
This NLSF error agrees with the shot-to-shot variation calculated as the root-mean square deviation (RMSD) of 10 spectra, each produced by averaging $N$ different shots. Only the RMSD for the 4.7~eV data is shown to declutter the graph.
The data is described well by $\Delta T_e \sim 1/\sqrt{N}$ and so the error scales inversely proportional to the SNR as expected.
An empirical formula based on scaling relations is derived from this data: equation \ref{formula} shows that  $\Delta T_e \sim \Delta \lambda_{1/e} \sim \sqrt{T_e}$. 
Simultaneously, the error decreases with SNR. SNR increases linearly with amplitude $A$ (counts) as $SNR \sim A \sim n_e/\Delta \lambda_{1/e} \sim n_e/\sqrt{T_e}$, since the area under the Gaussian is proportional to $A\Delta \lambda_{1/e}$. 
The formula 
\begin{equation}
    \Delta T_e = a\cdot (T_e/n_e) \cdot N^{-1/2}\ ,
\end{equation}   
with $a=8.2\times 10^{12}\ cm^{-3}$ describes all data in figure \ref{fig:error} well (orange curves), considering that the 1.2~eV plasma had a higher density (2.8$\times10^{13}$ cm$^{-3}$) than the 2.3~eV and 4.7~eV plasmas (1.2$\times 10^{13}$ cm$^{-3}$).

\section{Conclusion}
We have commissioned a TS system on the recently upgraded LAPD, and have performed the first measurements of electron density and temperature using the fully operational diagnostic.  
Measured densities and temperatures range from 4.0$\times$10$^{12}$ cm$^{-3}$  to 2.8$\times$10$^{13}$ cm$^{-3}$, and from 1.2~eV to 6.8~eV, depending on discharge parameters and in good agreement with Langmuir probe data. 
For a given pressure, the plasma temperature increases with the square of the discharge power. 
%For a constant bank voltage, a lower density results in higher temperature, consistent with a simple plasma model \cite{lieberman}.
The measured error $\Delta T_e$ is described well by an empirical formula derived from data.
In relatively colder plasma (1-2~eV) $T_e$ can be determined accurately by averaging as little as 10 discharges.

%\section{Data availability statement}
%The data that support the findings of this study are available from the corresponding author upon reasonable request.

\begin{acknowledgments}
This work was funded by the Department of Energy (DOE) under contract number DE-SC0021133. 
The Basic Plasma Science Facility (BaPSF) 
is supported by the National Science Foundation (PHY-1561912) and the DOE (DE-FC02-07ER54918). We thank Tai Ly and Marvin Drandell for their expert technical support.
\end{acknowledgments}

\bibliographystyle{unsrtnat}
\bibliography{bibliography.bib}

\end{document}